\documentclass{aastex631}

\usepackage{tikz}
\usetikzlibrary{shapes.geometric, arrows}
\usetikzlibrary{positioning}

\tikzset{
    startstop/.style={rectangle, rounded corners, minimum width=2.6cm,
        minimum height=1cm, text centered, draw=black, fill=gray!20},
    process/.style={rectangle, minimum width=3.2cm, minimum height=1cm,
        text centered, draw=black, fill=blue!20},
    decision/.style={diamond, aspect=2, text centered,
        draw=black, fill=green!20, inner sep=1pt},
    arrow/.style={thick,->,>=stealth}
}
\shorttitle{Common Origin of GRB X-ray Plateaus}
\shortauthors{Dong et al.}
\begin{document}

\title{Gamma-Ray Bursts: Evidence for a Common Origin of X-ray Plateaus with Diverse Temporal Decay Index}

\correspondingauthor{Yong-Feng Huang}
\email{hyf@nju.edu.cn}

\author[0009-0000-0467-0050]{Xiao-Fei Dong}
\affiliation{School of Astronomy and Space Science, Nanjing
University, Nanjing 210023, People's Republic of China}
\author[0000-0001-7199-2906]{Yong-Feng Huang}
\affiliation{School of Astronomy and Space Science, Nanjing
University, Nanjing 210023, People's Republic of China}
\affiliation{Key Laboratory of Modern Astronomy and Astrophysics
(Nanjing University), Ministry of Education, People's Republic of China}
\author[0000-0002-2191-7286]{Chen Deng}
\affiliation{School of Astronomy and Space Science, Nanjing University, Nanjing 210023, People's Republic of China}
\affiliation{Key Laboratory of Modern Astronomy and Astrophysics
(Nanjing University), Ministry of Education, People's Republic of China}
\author[0000-0002-6189-8307]{Ze-Cheng Zou}
\affiliation{School of Astronomy and Space Science, Nanjing University, Nanjing 210023, People's Republic of China}
\author[0000-0001-9648-7295]{Jin-Jun Geng}
\affiliation{Purple Mountain Observatory, Chinese Academy of Sciences, Nanjing 210023, People's Republic of China}
\author[0000-0001-7943-4685]{Fan Xu}
\affiliation{Institute of Space Weather, Nanjing University of Information Science and Technology, Nanjing 210023, People's Republic of China}
\author[0000-0002-5238-8997]{Chen-Ran Hu}
\affiliation{School of Astronomy and Space Science, Nanjing University, Nanjing 210023, People's Republic of China}
\author[0000-0003-3230-7587]{Orkash Amat}
\affiliation{School of Astronomy and Space Science, Nanjing University, Nanjing 210023, People's Republic of China}
\author[0009-0009-1579-5209]{Xiu-Juan Li}
\affiliation{School of Cyber Science and Engineering, Qufu Normal University, Qufu 273165, People's Republic of China}
\author[0000-0002-1343-3089]{Liang Li}
\affiliation{Institute of Fundamental Physics and Quantum Technology, Ningbo University, Ningbo, Zhejiang 315211, People's Republic of China}
\affiliation{School of Physical Science and Technology, Ningbo University, Ningbo, Zhejiang 315211, People's Republic of China}
\author[0000-0002-2162-0378]{Abdusattar Kurban}
\affiliation{State Key Laboratory of Radio Astronomy and Technology, Xinjiang Astronomical Observatory, CAS, 150 Science 1-Street, Urumqi, Xinjiang, 830011, People’s Republic of China}
\affiliation{Xinjiang Key Laboratory of Radio Astrophysics, Urumqi, Xinjiang, 830011, People’s Republic of China}

\begin{abstract}

A significant fraction of gamma-ray bursts (GRBs) exhibit a
plateau in the early X-ray afterglow light curve, whose mechanism
remains uncertain. While the post-plateau normal decay index
($\alpha_2$) is commonly used to constrain the afterglow dynamics,
the shallow-decay slope of the plateau itself ($\alpha_1$) has
received comparatively little attention. Recent observations,
however, reveal substantial dispersion in $\alpha_1$, raising the
question of whether GRBs with rising, flat and mildly decaying
plateaus represent intrinsically distinct populations. To address
this question, we collect a uniform sample of 185 \textit{Swift}
GRBs with a well-defined plateau and divide them into three groups
based on $\alpha_1$. Using a non-parametric approach, we
reconstruct their X-ray luminosity functions, redshift
distributions and event rates. It is found that the three groups
exhibit statistically consistent properties across all
diagnostics, with no evidence for group-specific features. Monte
Carlo perturbation tests further show that these results are
insensitive to the adopted classification boundaries of
$\alpha_1$. Our results indicate that variations in the plateau
slope $\alpha_1$ do not define distinct GRB subclasses, but
instead the sample constitutes a statistically uniform population
governed by a common framework.

\end{abstract}


\section{Introduction} \label{sec:intro}

The early X-ray afterglow of gamma-ray bursts (GRBs) is commonly
characterized by a canonical broken power-law temporal evolution,
comprising an initial steep decay, a shallow-decay (plateau), and
a subsequent normal decay \citep{2006Natur.444.1044G,
2006ApJ...642..354Z}. It is interesting to note that
\cite{2024ApJ...960...77D} recently analyzed 310
GRB X-ray afterglows and divided them into four temporal
categories based on the presence of flares, plateaus, and breaks.
While the steep decay is generally attributed to high-latitude
emission of the prompt phase \citep{2007ApJ...666.1002Z,
2025ApJ...989L..39Y} and the normal decay phase is broadly
consistent with the predictions of the standard external
forward-shock afterglow model \citep{1998ApJ...497L..17S,
1999MNRAS.309..513H}, the plateau phase -- characterized by an
unusually shallow temporal evolution over a timescale of
$\sim10^{2.5}$ -- $10^{4}$ s -- remains to be an enigmatic
component of the early afterglow \citep{2006MNRAS.366L..13G,
2007ApJ...670..565L}.

Owing to their well-defined temporal and luminosity properties,
X-ray plateaus provide a powerful diagnostic of the early
afterglow phase. They have been widely exploited to establish a
number of empirical correlations with potential cosmological
applications \citep{2008MNRAS.391L..79D, 2010ApJ...722L.215D,
2012A&A...538A.134X, 2019ApJS..245....1T, 2021ApJ...920..135X,
2023ApJ...943..126D}, to probe the features of external shock
emission and the circum-burst environment
\citep{2007ApJ...670..565L, 2012ApJ...744...36S,
2022ApJ...925...54T}, and to constrain long-lived central engine
activities \citep{2014MNRAS.443.1779R, 2015ApJ...805...89L,
2018ApJS..236...26L}.

The formation mechanism of the plateau remains a long-standing
question. Numerous models have been proposed, including continuous
energy injection from the central engine (e.g., a spinning
magnetar or a black hole undergoing fallback accretion) into the
blast wave \citep{2006MNRAS.369..197F, 2007ApJ...658L..75G,
2018ApJ...869..155S, 2025arXiv251111396D}; structured or off-axis
jets \citep{2018PhRvL.120x1103L, 2020A&A...641A..61A,
2022NatCo..13.5611D}; density stratification in the circum-burst
medium \citep{2006ApJ...640L.139T, 2007ApJ...656L..57J,
2025arXiv251207719Z}; reverse shock emission
\citep{2007MNRAS.381..732G, 2007ApJ...665L..93U,
2014MNRAS.442...20H, 2025MNRAS.tmp.1859F}; scattering echoes of
dust \citep{2008ApJ...675..507S}; and sub-relativistic cocoon or
shock-breakout scenarios \citep{2022ApJ...933..243F}. In addition,
an apparent plateau may sometimes arise from an inappropriate
choice of the reference time \citep{2025A&A...703A.101G}.

Given the diverse theoretical possibilities, it was naturally
expected that variations in the shallow-decay slope might reflect
different origins. For example, in the energy-injection
scenarios, sustained or effectively increasing power from a
newborn millisecond magnetar can produce a flat or rising plateau
\citep{Dai1998, 2001ApJ...552L..35Z}, whereas the declining
fallback accretion onto a nascent black hole may result in a
mildly decaying plateau \citep{2015MNRAS.446.3642Y,
2018ApJ...857...95M}. By contrast, structured jets and
viewing-angle effects can produce diverse plateau morphology
without requiring special central-engine evolution
\citep{2002MNRAS.332..945R, 2003ApJ...591.1086G,
2020MNRAS.492.2847B}.

However, different scenarios can produce similar plateau slopes,
leading to significant degeneracy in the observed light curves.
Consequently, many authors have relied on the post-plateau decay
to infer or constrain the plateau origin, rather than directly
exploiting the plateau slope itself. A normal post-plateau decay
index of order unity is generally consistent with external-shock
energy injection \citep{2007ApJ...662.1093W}, whereas an extremely
steep decay ($\lesssim -3$) is commonly interpreted as an
``internal plateau'' connected to abruptly terminated
central-engine activities \citep{2007ApJ...665..599T,
2010MNRAS.402..705L, 2017ApJ...849..119C, 2017A&A...605A..60B}.

Recent analyses have shown that the shallow-decay index displays
non-negligible dispersion among different bursts
\citep{2019ApJS..245....1T, 2023ApJ...943..126D,
2023A&A...675A.117R, 2024A&A...692A..73G}, suggesting potential
diversities in the underlying mechanisms. Here we conduct a
systematic population-level study to test whether the distribution
of the shallow-decay index provides additional constraints on the
origin of the X-ray plateau.

The paper is organized as follows. Section~\ref{sec:Data}
describes the data set, subsample classification, and parameter
calculations. Section~\ref{sec:method} outlines the non-parametric
methods used to derive the intrinsic luminosity function and event
rate. The results for each subsample are presented in
Section~\ref{sec:results}. In Section~\ref{sec:mctest}, we further
assess the robustness of our results via Monte Carlo simulations.
Section~\ref{sec:conclusion} summarizes the main findings, and
Section~\ref{sec:Discussion} discusses the key implications.

\section{Data} \label{sec:Data}
\subsection{Sample Selection}
\label{sec:sample}
To investigate whether variations in the plateau decay index
reflect different GRB populations, we need a large, uniformly
selected sample with well defined X-ray plateaus. The X-Ray
Telescope (XRT) onboard \textit{Neil Gehrels Swift Observatory}
has detected over 1700 GRBs \citep{2004ApJ...611.1005G}, and the
corresponding X-ray light curves and spectra are publicly
available through the \textit{Swift}-XRT
Repository\footnote{\url{https://www.swift.ac.uk/xrt_curves/}}
\citep{2007A&A...469..379E, 2009MNRAS.397.1177E}.

\cite{2019ApJS..245....1T} systematically analyzed all
\textit{Swift}/XRT GRBs observed between March 2005 and August
2018, aiming to study potential empirical relations involving
plateau parameters. Their sample selection followed four criteria:
(i) the temporal index of the plateau should be in the range
$-1.0$ to $+1.0$; (ii) adequate data are available to define the
plateau clearly; (iii) no flares were observed during the plateau
phase; and (iv) the redshift should be available. A total of 174
GRBs (including 7 short bursts) satisfied these requirements. Each
light curve was fitted with the smoothly broken power-law function
\citep{2007ApJ...670..565L, 2012ApJ...758...27L,
2016ApJS..224...20Y}
\begin{equation}
\label{func1}
F_{\rm X}(t) =  F_{\rm X,0} \left[
\left( \frac{t}{T_{\rm a,obs}} \right)^{\alpha_{1}\omega} +
\left( \frac{t}{T_{\rm a,obs}} \right)^{\alpha_{2}\omega}
\right]^{-1/\omega},
\end{equation}
where $\alpha_1$ and $\alpha_2$ denote the temporal slopes before
and after the break, $T_{\rm a,obs}$ is the observed break time,
$\omega$ sets the smoothness of the transition, and $F_{\rm X,0}
\cdot 2^{-1/\omega}$ corresponds to the flux at $T_{\rm a,obs}$.
The best-fit parameters were obtained via a Markov Chain Monte
Carlo (MCMC) procedure.

Following a similar methodology, \cite{2023ApJ...943..126D}
reanalyzed all \textit{Swift}/XRT GRBs detected between March 2005
and May 2022, fixing the smoothness parameter to $\omega = 1$. To
expand the sample size as far as possible, they included GRBs
partially overlapped by X-ray flares. The flare intervals were
excluded during the fitting procedure. This treatment is
reasonable in light of more recent results (e.g.,
\citealt{2025arXiv251207731D}), which indicate that X-ray flares
originate from a process different from the plateau emission. As a
result, they added 36 more GRBs with measured redshifts into the
plateau sample, including 2 short bursts (GRB~090426 and
GRB~100724A).

In this study, we further examined all \textit{Swift}/XRT GRBs
observed between June 2022 and August 2025 based on data from the
\textit{Swift}/XRT Repository, following the same selection
criteria as \cite{2023ApJ...943..126D}. This yielded 17 new GRBs
with measured redshifts. Two short bursts (GRB~231117A and
GRB~250221A) were excluded since we mainly concentrate on long
events. The remaining 15 long GRBs were fitted with
Eq.~(\ref{func1}) using an MCMC method, and their best-fit
parameters are listed in Table~\ref{tab1}.

Given the systematic differences between long and short GRBs
\citep{2013MNRAS.428..729M} and the small number of short bursts,
we restricted our study only to long GRBs. In total, we collected
216 long GRBs (167 from \cite{2019ApJS..245....1T}, 34 from
\cite{2023ApJ...943..126D}, and 15 newly added here). To give a
more strict definition for the plateau phase, here we require that
the temporal index should be in the range of $-$0.5 -- $+$0.5.
This yields 185 GRBs, which constitute our final plateau sample.

\begin{table}[htbp]
    \centering
    \caption{Key parameters of the 15 long $Swift$ GRBs with an X-ray Plateau (after June 2022).}
    \label{tab1}
    \begin{tabular}{lccccccccc}
        \hline
        GRB name & $z$ & $ T_{\rm 90}$ & ${\rm log} (F_{\rm X,0}/10^{-12})$ & ${\rm log} (T_{\rm a}/10^{3})$ &
        $\alpha_{\rm 1}$ & $\alpha_{\rm 2}$ & $\Gamma$ & ${\rm log} (L_{\rm X}/10^{47})$ & \tablenotemark{\rm *}Reference of z \\
         &  & (s) & (${\rm erg\ cm^{-2}\ s^{-1}}$) & (s) &  &  &  & (${\rm erg\ s^{-1}}$) &  \\
        (1)& (2) & (3) & (4) & (5) & (6) & (7) & (8) & (9) & (10) \\
        \hline
    220611A & $2.36$ & $57$ & $0.35^{+0.05}_{-0.03}$ & $1.97^{+0.03}_{-0.06}$ & $0.23^{+0.03}_{-0.03}$ & $1.55^{+0.07}_{-0.08}$ & $1.9$ & $-0.35^{+0.05}_{-0.03}$ & $1$ \\
    220813A & $0.82$ & $\sim 30$\tablenotemark{b} & $1.13^{+0.25}_{-0.34}$ & $0.53^{+0.42}_{-0.37}$ & $0.26^{+0.18}_{-0.28}$ & $1.16^{+0.12}_{-0.1}$ & $2$ & $-0.66^{+0.25}_{-0.34}$ & $2$ \\
    230116D & $3.81$ & $41$ & $0.87^{+0.36}_{-0.33}$ & $0.85^{+0.26}_{-0.34}$ & $0.34^{+0.22}_{-0.4}$ & $2.09^{+0.61}_{-0.36}$ & $1.76$ & $0.56^{+0.36}_{-0.33}$ & $3$ \\
    230414B & $3.57$ & $25.98$ & $0.4^{+0.11}_{-0.09}$ & $1.73^{+0.07}_{-0.08}$ & $0.00^{+0.13}_{-0.16}$ & $2.9^{+0.45}_{-0.37}$ & $2.02$ & $0.2^{+0.11}_{-0.11}$ & $4$ \\
    230818A & $2.42$ & $9.82$ & $0.67^{+0.21}_{-0.14}$ & $1.14^{+0.12}_{-0.2}$ & $0.41^{+0.08}_{-0.13}$ & $1.97^{+0.24}_{-0.22}$ & $2$ & $0.05^{+0.21}_{-0.14}$ & $5$ \\
    240419A & $5.18$ & $3$ & $1.17^{+0.15}_{-0.25}$ & $0.24^{+0.28}_{-0.23}$ & $-0.07^{+0.3}_{-0.35}$ & $1.4^{+0.2}_{-0.17}$ & $2.18$ & $1.48^{+0.15}_{-0.25}$ & $6$ \\
    240529A & $2.7$ & $160.67$ & $1.91^{+0.03}_{-0.03}$ & $1.27^{+0.02}_{-0.02}$ & $0.11^{+0.04}_{-0.04}$ & $2.29^{+0.06}_{-0.06}$ & $2.14$ & $1.48^{+0.03}_{-0.03}$ & $7$ \\
    241010A & $0.98$ & $30.86$ & $1.56^{+0.08}_{-0.08}$ & $1.06^{+0.09}_{-0.09}$ & $0.18^{+0.04}_{-0.05}$ & $1.53^{+0.06}_{-0.06}$ & $1.91$ & $-0.06^{+0.08}_{-0.08}$ & $8$ \\
    241026A & $2.79$ & $25.2$ & $1.21^{+0.08}_{-0.08}$ & $1.58^{+0.07}_{-0.07}$ & $0.3^{+0.03}_{-0.04}$ & $2.07^{+0.13}_{-0.12}$ & $1.86$ & $0.66^{+0.08}_{-0.08}$ & $9,10$ \\
    250101A & $2.49$ & $34.19$ & $1.07^{+0.25}_{-0.15}$ & $0.79^{+0.16}_{-0.28}$ & $0.34^{+0.07}_{-0.15}$ & $1.48^{+0.12}_{-0.12}$ & $2.04$ & $0.5^{+0.25}_{-0.15}$ & $11,12$ \\
    250108B & $2.2$ & $229.66$ & $0.81^{+0.07}_{-0.27}$ & $1.06^{+0.49}_{-0.19}$ & $-0.22^{+0.43}_{-0.45}$ & $0.95^{+0.12}_{-0.07}$ & $2.07$ & $0.12^{+0.07}_{-0.27}$ & $13$ \\
    250114A & $4.73$ & $293.89$ & $0.27^{+0.13}_{-0.12}$ & $1.5^{+0.08}_{-0.09}$ & $0.5^{+0.1}_{-0.13}$ & $3.58^{+0.51}_{-0.48}$ & $1.85$ & $0.23^{+0.13}_{-0.12}$ & $14$ \\
    250129A & $2.15$ & $262.25$ & $1.16^{+0.08}_{-0.1}$ & $0.97^{+0.13}_{-0.12}$ & $-0.12^{+0.14}_{-0.15}$ & $1.38^{+0.1}_{-0.08}$ & $1.94$ & $0.39^{+0.08}_{-0.1}$ & $15,16$ \\
    250424A & $0.31$ & $19.03$ & $2.14^{+0.06}_{-0.07}$ & $0.78^{+0.1}_{-0.09}$ & $0.05^{+0.06}_{-0.07}$ & $1.16^{+0.03}_{-0.03}$ & $2.07$ & $-0.65^{+0.06}_{-0.07}$ & $17$ \\
    250430A & $0.77$ & $9.17$ & $1.68^{+0.13}_{-0.29}$ & $0.02^{+0.33}_{-0.22}$ & $-0.27^{+0.47}_{-0.46}$ & $1.73^{+0.93}_{-0.68}$ & $1.89$ & $-0.2^{+0.13}_{-0.29}$ & $18$ \\
    \hline
    \end{tabular}
    \vspace{0.5em}
    \begin{minipage}{0.95\linewidth}
        \small
        \textbf{Note.} --
        Column (1): GRB name;
        (2): redshift;
        (3): duration from $Swift$/BAT in the 15 -- 150 keV energy band; \tablenotemark{b}GRB 220813A showed a single-peaked structure with a duration of about 30 s \citep{2022GCN.32465....1B};
        (4): the flux at the end of the plateau is $F_{\rm X,0}/2$;
        (5): the break time in the observer's frame;
        (6): temporal power-law decay index of the plateau phase;
        (7): temporal power-law decay index of the post break segment;
        (8): photon index of the X-ray spectrum;
        (9): the isotropic X-ray luminosity at the break time after the $K$-correction;
        (10): references for the redshift measurement. \tablenotetext{*}{
    1.\cite{2022GCN.32595....1S};
    2.\cite{2022GCN.32471....1F};
    3.\cite{2023GCN.33187....1M};
    4.\cite{2023GCN.33629....1A};
    5.\cite{2023GCN.34485....1M};
    6.\cite{2024GCN.36176....1S};
    7.\cite{2024GCN.36574....1D};
    8.\cite{2024GCN.38080....1M};
    9.\cite{2024GCN.37916....1M};
    10.\cite{2024GCN.37925....1I};
    11.\cite{2025GCN.38759....1Z};
    12.\cite{2025GCN.38776....1L};
    13.\cite{2025GCN.38877....1M};
    14.\cite{2025GCN.38934....1M};
    15.\cite{2025GCN.39071....1S};
    16.\cite{2025GCN.39073....1I};
    17.\cite{2025GCN.40228....1S};
    18.\cite{2025GCN.40301....1G}
    }
    \end{minipage}
\end{table}

\subsection{Sub-samples and Statistical Properties}
\begin{figure*}[ht!]
\centering
\includegraphics[width=0.97\textwidth, trim=0 0 0 0, clip]{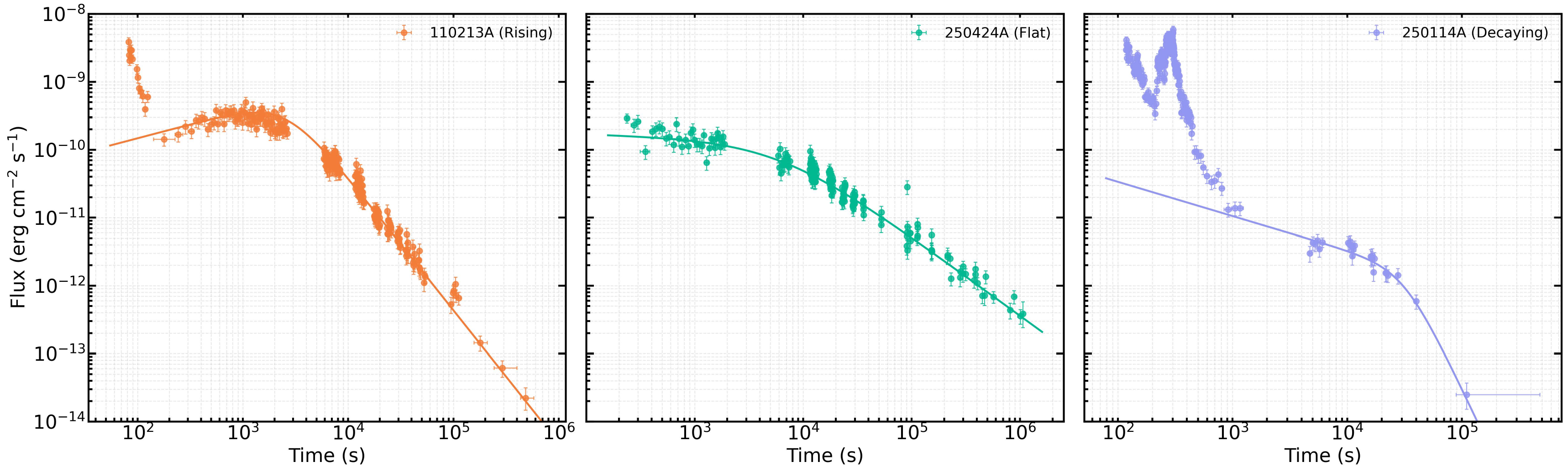}
\caption{Exemplar X-ray afterglow light curves of GRBs~110213A,
250424A, and 250114A, which has a rising, flat,
and decaying plateau, respectively. The solid curves show
the best-fit results by using Equation~(\ref{func1}) with the MCMC
method. } \label{fig1}
\end{figure*}

We notice that the X-ray luminosity is not strictly constant
during the plateau phase. It can either decrease or increase
slightly, which is reflected in the variation of the plateau
temporal index $\alpha_1$. To investigate whether such a behavior
is due to different mechanisms, we divide the 185 long GRBs into
three subsamples by considering the value of $\alpha_1$: bursts
with $-0.5 < \alpha_1 \leq -0.1$ are called the \textit{Rising}
group, those with $-0.1 < \alpha_1 \leq 0.1$ are the \textit{Flat}
group, and those with $0.1 < \alpha_1 \leq 0.5$ are the
\textit{Decaying} group. The three subsamples have a rising, flat,
or decaying plateau, respectively. Figure~\ref{fig1} shows a
representative X-ray light curve of each group, together with the
best-fit curve. In this way, the \textit{Rising}, \textit{Flat},
and \textit{Decaying} subsamples contain 16, 75, and 94 GRBs,
respectively.

\begin{figure*}[ht!]
\centering
\includegraphics[width=0.97\textwidth, trim=0 0 0 0, clip]{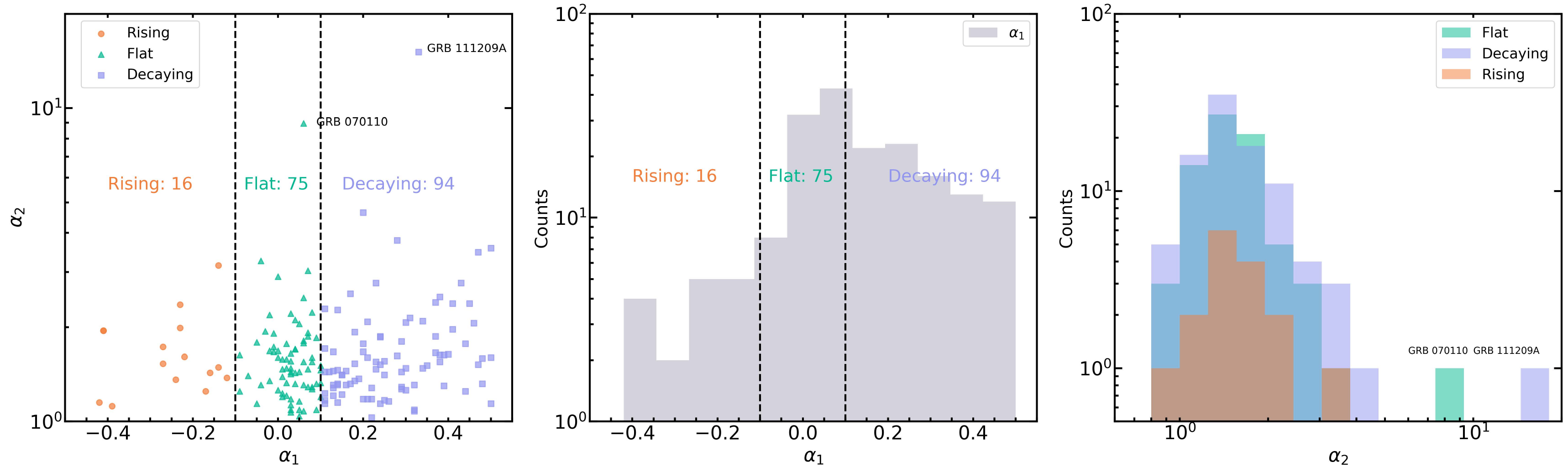}
\caption{
Distributions of the temporal indices during ($\alpha_1$) and after ($\alpha_2$) the plateau phase for the three groups.
 \textbf{Left:} $\alpha_2$ plotted versus $\alpha_1$, where
 the filled circles, triangles and squares represent the
 \textit{Rising}, \textit{Flat} and \textit{Decaying} subsamples, respectively.
 \textbf{Middle:} histogram of $\alpha_1$ for all the plateau GRBs.
 \textbf{Right:} histograms of $\alpha_2$ for the three groups.
}
\label{fig2}
\end{figure*}

The overall distribution of $\alpha_1$ and $\alpha_2$ is shown in
Figure~\ref{fig2}. The left panel plots $\alpha_1$ versus
$\alpha_2$. We see that these two indices are generally
independent of each other. In fact, we have explored any potential
correlation between $\alpha_1$ and $\alpha_2$ and obtained a
Pearson coefficient of $r = 0.11$, with a significance of $p
\simeq 0.13$, indicating a very weak and statistically
insignificant correlation. In the middle panel, the distribution
of $\alpha_1$ is not symmetrical, but slightly skewed toward
positive values, peaking at $\alpha_1 \simeq 0.1$, consistent with
the results of \cite{2019ApJS..245....1T}. In the right panel, we
see that $\alpha_2$ lies mainly in $0.8 \lesssim \alpha_2 \lesssim
4.0$. There are no clear difference in the distribution of
$\alpha_2$ for the three subsamples.

A small number of GRBs display an exceptionally steep decay rate
after the plateau, including GRB~070110 ($\alpha_2 \approx 8.95$),
GRB~100219A ($\alpha_2 \approx 4.64$), GRB~100902A ($\alpha_2
\approx 4.69$), GRB~111209A ($\alpha_2 \approx 15.11$) and
GRB~170714A ($\alpha_2 \approx 4.97$). These rapid declines are
typical of internal plateau bursts, potentially due to the sudden
termination of the central-engine activity
\citep{2007ApJ...665..599T, 2010MNRAS.402..705L,
2017ApJ...849..119C, 2017A&A...605A..60B}. Figure~\ref{fig2} also
highlights two representative examples -- GRB~070110 and
GRB~111209A -- to illustrate their extreme $\alpha_2$ values.

We have calculated the difference between $\alpha_1$ and
$\alpha_2$, i.e. $\Delta \alpha = \alpha_2 - \alpha_1$. It is
found that the average difference is $\langle \Delta \alpha
\rangle = 1.598$, which is consistent with earlier results of
\cite{2007ApJ...670..565L} ($1.11 \pm 0.39$) and
\cite{2019ApJS..245....1T} ($1.4 \pm 0.3$). Given the relatively
narrow distribution of $\alpha_2$ and its weak correlation with
$\alpha_1$, classifying GRBs into \textit{Rising}, \textit{Flat},
and \textit{Decaying} subsamples based on $\alpha_1$ alone is
reasonable.

\subsection{Plateau Luminosity}
\label{sec:Data Analysis}

Based on the fitted ${F_{\rm X,0}}$ parameter, the isotropic X-ray
luminosity at the end of the plateau can be calculated as
\begin{eqnarray}
L_{\rm X} = 4\pi D_{\rm L}^{2}(z)\,\frac{F_{\rm X,0}}{2}\,K,
\end{eqnarray}
where $K$ is the $K$-correction factor, and $D_{\rm L}(z)$ is the
luminosity distance,
\begin{eqnarray}
D_{\rm L}(z) = (1+z)\frac{c}{H_{0}}\int_{0}^{z}\frac{dz^{\prime}}{\sqrt{\Omega_{\rm M}(1+z^{\prime})^{3}+\Omega_{\Lambda}}},
\end{eqnarray}
with $c$ being the speed of light. Here, we adopt a flat Lambda 
Cold Dark Matter ($\Lambda$CDM) cosmology with 
$H_0 = 67.3~{\rm km~s^{-1}~Mpc^{-1}}$,
$\Omega_{\rm M} = 0.315$, and $\Omega_\Lambda = 1-\Omega_{\rm M}$
\citep{2014A&A...571A..16P, 2020A&A...641A...6P}.

The $K$-correction \citep{2001AJ....121.2879B} is calculated as
\begin{eqnarray}
K = \frac{\int_{E_{1}/(1+z)}^{E_{2}/(1+z)} E\,N(E)\,dE}{\int_{E_{1}}^{E_{2}} E\,N(E)\,dE},
\end{eqnarray}
where $(E_1, E_2)$ denotes the energy band of the detector.
Considering the relatively narrow energy coverage of
\textit{Swift}/XRT, we approximate the photon spectrum by a simple
power law, $N(E) = N_0 E^{-\Gamma}$, where $\Gamma$ is the photon
index measured by \textit{Swift}/XRT. The values of $\Gamma$ are
taken from the \textit{Swift} GRB online
catalogue\footnote{\url{https://swift.gsfc.nasa.gov/archive/grb_table/}}.
With this power-law spectrum, the isotropic X-ray luminosity
simplifies to
\begin{eqnarray}
L_{\rm X} = \frac{4\pi D_{\rm
L}^{2}(z)}{(1+z)^{2-\Gamma}}\,\frac{F_{\rm X,0}}{2},
\end{eqnarray}
after considering the $K$-correction.

\begin{figure*}[ht!]
\centering
\includegraphics[width=0.97\textwidth, trim=0 0 0 0, clip]{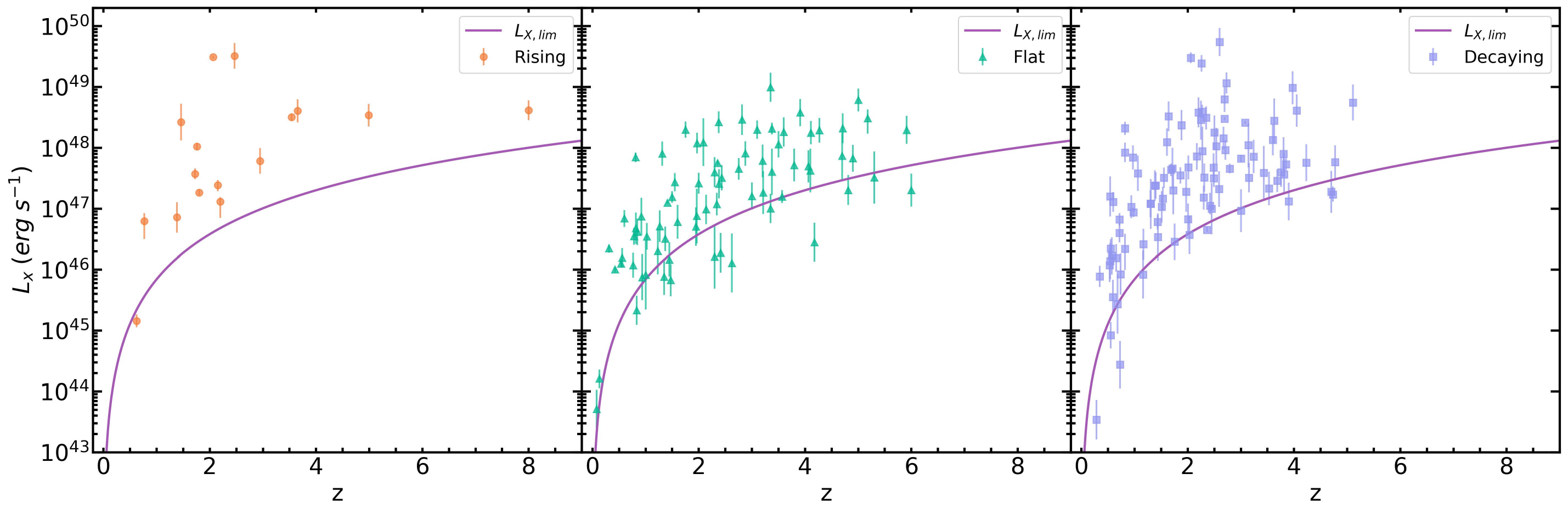}
\caption{X-ray plateau luminosity plotted versus redshift for the
three subsamples. The solid curves are plotted by assuming a flux
limit of $\log(F_{\rm lim}/{\rm 1\ erg\ cm^{-2}\ s^{-1}} )=-11.9$
\citep{2025ApJ...990...69K}, which will be taken as the detection
threshold to ensure sample completeness in our subsequent
calculations. } \label{fig3}
\end{figure*}

In Table~\ref{tab1}, we have also listed the derived plateau
luminosity for the 15 new long GRBs observed by \textit{Swift}/BAT
between June 2022 and August 2025. For the other earlier long GRBs
with X-ray plateaus, the isotropic luminosity data are directly
taken from \citet{2019ApJS..245....1T} and
\citet{2023ApJ...943..126D}. The distribution of $L_{\rm X}$ is
shown in Figure~\ref{fig3} for the three subsamples. Following
\cite{2025ApJ...990...69K}, we have assumed a flux threshold of
$\log(F_{\rm lim}/{\rm 1\ erg\ cm^{-2}\ s^{-1}})=-11.9$ to ensure
sample completeness in our subsequent analysis. The derived
limiting curve is also shown in Figure~\ref{fig3}. We see that
nearly all rising-plateau GRBs are located above the detection
threshold, while a small number of events in the flat and decaying
subsamples are below the threshold. Such a distribution hints that
the rising plateaus preferentially occur in systems with a
stronger engine power.

\section{Modeling the Event Rate}
\label{sec:method}

We have grouped the 185 long GRBs with
well-defined X-ray plateaus into three subsamples according to
their temporal index $\alpha_{1}$. In this section, we compare
these subsamples in terms of their luminosity function,
redshift distribution, and event rate to examine whether the various plateau
behaviors are due to different intrinsic physical processes.

To derive the intrinsic luminosity function and event rate,
corrections for redshift evolution and observational biases are
required. Malmquist bias and instrumental flux truncation can
induce artificial luminosity evolution and selection effects. To
mitigate these, we employ the non-parametric Lynden-Bell $C^{-}$
method \citep{1971MNRAS.155...95L}, which provides an unbiased
estimate of the intrinsic event rate distribution in flux-limited
samples.

The $C^{-}$ estimator is a non-parametric maximum-likelihood
approach developed for randomly truncated data and has been widely
adopted for GRB event rate estimation \citep{2015ApJS..218...13Y,
2016A&A...587A..40P, 2018ApJ...852....1Z, 2021RAA....21..254L,
2022MNRAS.513.1078D, 2023ApJ...958...37D, 2025ApJ...993...20D,
2025arXiv251023945C}. It is also effective in correcting for
selection biases and parameter evolution
\citep{1992ApJ...399..345E, 2013ApJ...774..157D,
2017A&A...600A..98D, 2021ApJ...920..135X} and underpinning more
advanced statistical inferences such as two-sample test,
correlation, linear regression, and Bayesian analysis
\citep{1992ApJ...399..345E, 2012msma.book.....F, Dorre&Emura2019}.

Before applying the $C^{-}$ method, the truncation boundary
defined by the detector sensitivity must be specified.
\citet{2025ApJ...990...69K} followed the procedure outlined in
\citet{2021ApJ...914L..40D}, which examine the completeness of their
plateau GRB dataset to select the optimal flux threshold value.
They found that different flux thresholds yield broadly 
similar GRB event rates, a result that was also confirmed by 
\citet{2022MNRAS.513.1078D} and \citet{2025MNRAS.542..215L}. 
In this study , we adopt a flux limit of 
$\log(F_{\rm lim}/{\rm 1\ erg\ cm^{-2}\ s^{-1}})=-11.9$, 
following their choice. The sensitivity
curve is then calculated as $L_{\rm X,lim} = 4\pi D_{\rm
L}^{2}(z)F_{\rm lim}/(1+z)^{2-\Gamma_{\rm avg}}$, as shown by the
solid line in Figure~\ref{fig3}, where $\Gamma_{\rm
avg}\simeq1.99$ is the average photon index.

An additional requirement of the $C^{-}$ method is that the
luminosity ($L_{\rm X}$) and redshift ($z$) should be
statistically independent, implying no intrinsic luminosity
evolution. We test and remove such evolution using the
non-parametric Efron-Petrosian (EP) method
\citep{1992ApJ...399..345E}. After the de-evolved data set is
ready, the $C^{-}$ method can be applied to reconstruct the
intrinsic luminosity and redshift distributions with various
biases being removed as far as possible
\citep{2013ApJ...774..157D}.

\subsection{EP Method}
The EP method (also known as the $\tau$-statistic test;
\citealt{1992ApJ...399..345E}) is employed to remove the redshift
evolution of luminosity, i.e., the intrinsic correlation between
$L_{\rm X}$ and $z$. Following \cite{2025ApJ...990...69K}, we
assume that the observed luminosity evolves with redshift as
$L_{\rm X} \propto (1+z)^{k}$, where $k$ is a constant quantifying
the degree of evolution. Once $k$ is determined, the
evolution-corrected luminosity can be expressed as $L'_{\rm X} =
L_{\rm X}/(1+z)^{k}$. This correction yields an independent
parameter pair $(z, L'_{\rm X})$, such that the GRB distribution
function can be factorized as 
$\Psi(L'_{\rm X}, z) = \psi(L'_{\rm X})\,\phi(z)$, 
where $\psi(L'_{\rm X})$ and $\phi(z)$ represent the luminosity function
and redshift distribution, respectively.

To estimate $k$, each GRB data point $(z_i, L_{\rm X,i})$
is transformed into $(z_i, L'_{\rm X,i})$ for a given trial $k$.
For the $i$th GRB, we define a comparable data set
\begin{equation}\label{Ji}
J_i = \{\,j \mid L'_{\rm X,j} \ge L'_{\rm X,i},~ z_j \le z_i^{\rm max}\,\},
\end{equation}
where $z_i^{\rm max}$ is the maximum redshift at which a burst with
luminosity $L'_{\rm X,i}$ can be detected by \textit{Swift}/XRT.
The number of GRBs in this region is denoted by $n_i$,
and the number of sources with $z \le z_i$ is $R_i$.
The $\tau$ statistic is defined as
\begin{equation}
\tau = \frac{\sum_i (R_i - T_i)}{\sqrt{\sum_i V_i}},
\end{equation}
where $T_i = (1 + n_i)/2$ and $V_i = (n_i^2 - 1)/12$
are the expected mean and variance of $R_i$, respectively.

If luminosity and redshift are independent, $R_i$ should be uniformly
distributed between 1 and $n_i$, giving $\tau \approx 0$.
Otherwise, the trial value of $k$ is adjusted and the process repeated
until $\tau = 0$ is achieved, yielding the best-fit evolution index.
Using this procedure, we find $k \simeq 5.53$ for the rising
sample, $k \simeq 4.17$ for the flat sample, and $k \simeq 4.32$
for the decaying sample. The X-ray plateau luminosities are then
corrected by dividing them by $(1+z)^{k}$.

\subsection{Lynden--Bell's $C^{-}$ Method}

The Lynden--Bell's $C^{-}$ method offers an effective approach 
for recovering the intrinsic bivariate distributions of 
astronomical sources from truncated data sets.
For each GRB, we define a comparable data set
\begin{equation}
J'_i = \{\,j \mid L'_{\rm X,j} \ge L_{\rm X,i}^{'{\rm min}},~ z_j < z_i \,\},
\end{equation}
where $L_{\rm X,i}^{'{\rm min}}$ is the
minimum detectable luminosity
at $z_i$, and $M_i$ denotes the number of
GRBs contained in $J'_i$.

According to the $C^{-}$ formalism,
the cumulative luminosity function of GRB plateaus can be estimated as
\begin{equation}\label{luminosityFunction}
\psi(L'_{\rm X,i}) = \prod_{j>i} \left(1+\frac{1}{N_j}\right),
\end{equation}
where the product runs over all GRBs with
de-evolved luminosities $L'_{\rm X,j} > L'_{\rm X,i}$.
Similarly, the cumulative redshift distribution is given by
\begin{equation}\label{redshiftFunction}
\phi(z_i) = \prod_{j<i} \left(1+\frac{1}{M_j}\right),
\end{equation}
where the product runs over all GRBs with $z_j < z_i$.
The comoving event rate of GRBs is then obtained as
\begin{equation}\label{formationrate}
\rho(z) = (1+z)\,\frac{d\phi(z)}{dz}\,
\left[\frac{dV(z)}{dz}\right]^{-1},
\end{equation}
where the factor $(1+z)$ accounts for cosmological time dilation,
and $dV(z)/dz$ is the differential comoving volume element
\citep{2019JHEAp..24....1K}:
\begin{equation}\label{comovingvolume}
\frac{dV(z)}{dz} =
\frac{c}{H_0}\,
\frac{4\pi D_{\rm L}^2(z)}{(1+z)^2}\,
\frac{1}{\sqrt{(1 - \Omega_{\rm M}) + \Omega_{\rm M}(1+z)^3}}.
\end{equation}

\section{Results} \label{sec:results}

In this section, we present our main results obtained for
the rising, flat, and decaying subsamples.
First, we compare the luminosity functions of the three groups.
We then derive their redshift-dependent event rates
and compare them with the cosmic star-formation rate (SFR)
to explore potential links between the plateau GRBs
and star-forming environments.

\subsection{Luminosity Functions}

\begin{figure*}[ht!]
\centering
\includegraphics[width=0.55\textwidth, trim=0 0 0 0, clip]{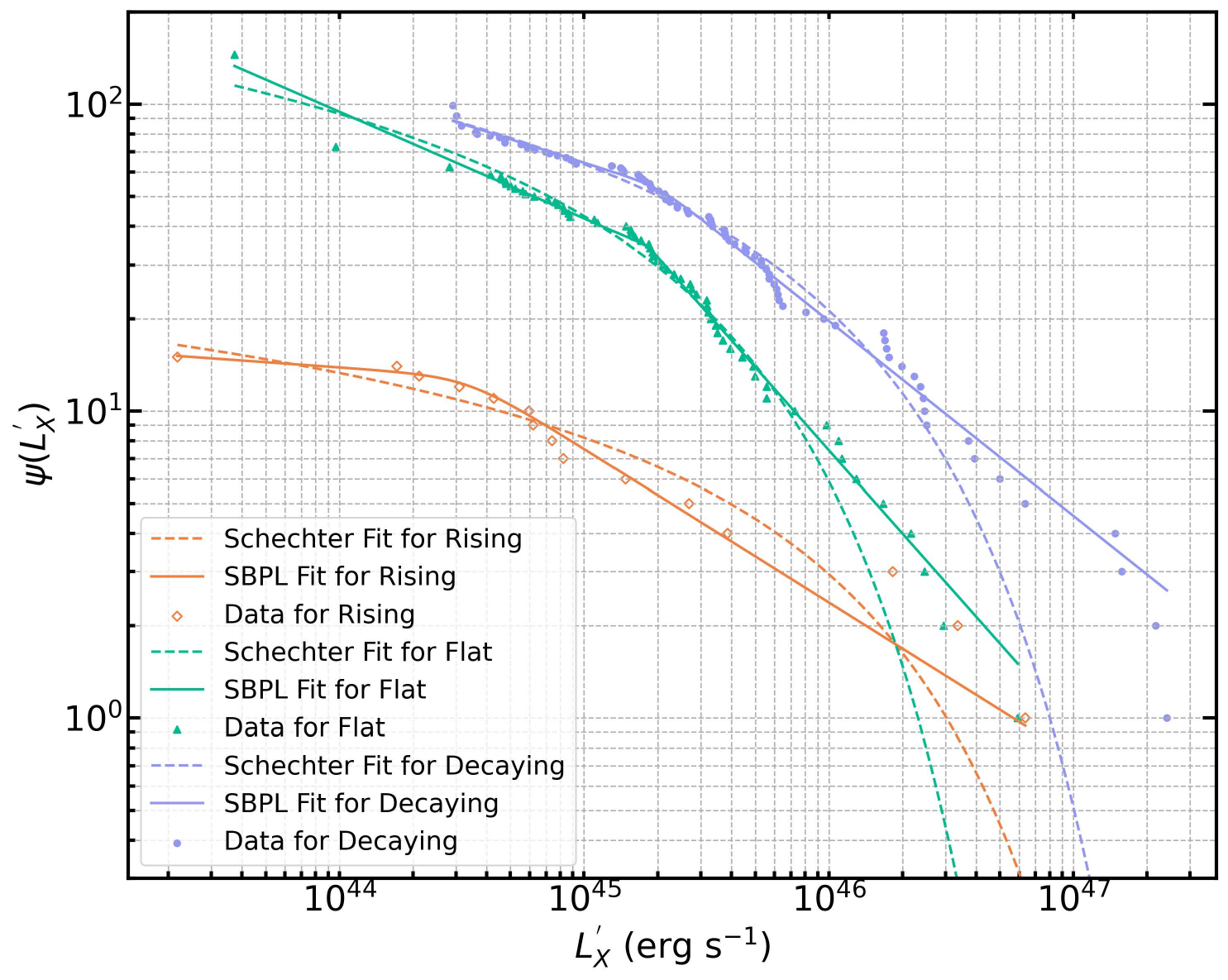}
\caption{ Luminosity functions of the three subsamples. The data
points have been calculated by using
Equation~(\ref{luminosityFunction}), where the diamonds, triangles
and filled circles represent the rising, flat and decaying
subsamples, respectively. The dashed line denote the best-fit
Schechter function for each subsample, while the solid curve shows
the best-fit SBPL model. The corresponding best-fit parameters are
listed in Table~\ref{tab2}. } \label{fig4}
\end{figure*}

Figure \ref{fig4} presents the luminosity functions of the three
subsamples, computed using Equation (\ref{luminosityFunction}).
Generally, all the three curves exhibit similar shapes, declining
as the luminosity increases, with a shallow slope at low
luminosity segment and a steeper drop at the high luminosity
regime, with only minor fluctuations. The rising and flat
subsamples extend to slightly lower luminosity segment than the
decaying one, but the reasons are different. In the flat
subsample, two nearby low-luminosity bursts pass the flux
threshold (GRB 051109B at $z \approx 0.08$ and GRB 060614 at $z
\approx 0.13$; \citealt{2006GCN..5387....1P, 2006Natur.444.1044G,
2007A&A...470..105M}), whereas the rising sample reaches a lower
luminosity mainly because a stronger redshift-luminosity evolution
correction (i.e., a larger $k$) is applied.

\begin{table}[htbp]
    \centering
    \caption{Best-fit parameters of the Schechter function and the smoothly broken power-law function.}
    \label{tab2}
    \begin{tabular}{lcccc}
        \hline
        \multicolumn{5}{c}{\textbf{Schechter Function}}\\
        \hline
        Sample & $\psi_\star$ & $L_\star$ ($10^{46}$ $\rm erg\ s^{-1}$) & $\gamma$ & $R^{2}$ \\
        (1) & (2) & (3) & (4) & (5) \\
        \hline
        Rising & $2.82\pm1.20$ & $4.12\pm3.42$ & $-1.95\pm0.10$ & $0.9284$ \\
        Flat & $22.4\pm4.25$ & $1.15\pm0.27$ & $-2.00\pm0.06$ & $0.9467$ \\
        Decaying & $20.3\pm2.47$ & $4.01\pm0.65$ &$ -2.00\pm0.03$ & $0.9859$  \\
        \hline
        \multicolumn{5}{c}{\textbf{Smoothly Broken Power-law Function}}\\
        \hline
        Sample & $L_{\rm break}$  ($10^{45}$ $\rm erg\ s^{-1}$) & $A_1$ & $A_2$ & $R^{2}$ \\
        (1) & (2) & (3) & (4) & (5) \\
        \hline
        Rising & $0.34\pm1.19$ & $0.06\pm0.12$ & $0.5\pm0.03 $& $0.9810$\\
        Flat & $1.83\pm0.17$ & $0.35\pm0.01$ & $0.90\pm0.09$ & $0.9769$  \\
        Decaying & $2.06\pm0.15$ & $0.25\pm0.01$ & $0.64\pm0.02 $& $0.9930 $ \\
        \hline
    \end{tabular}
\end{table}

To better understand the luminosity function of each subsample, we
have tried to fit it with both a Schechter model,
\begin{equation}
\psi(L) = \psi_\star \left(\frac{L}{L_\star}\right)^{\gamma}
\exp\left(-\frac{L}{L_\star}\right),
\end{equation}
and a smoothly broken power-law (SBPL) model,
\begin{equation}
\psi(L) = \psi_0 \left[\left(\frac{L}{L_{\rm break}}\right)^{A_1\, \rm w}
+ \left(\frac{L}{L_{\rm break}}\right)^{A_2 \, \rm w}\right]^{-1/\rm w},
\end{equation}
where ${L_\star}$ and $L_{\rm break}$ are the characteristic
luminosity and ${\gamma}$, $A_1$ and $A_2$ denote the slopes. We
then compare the goodness of fit of these two models.

The best-fit curves of the two kinds of models are shown in
Figure~\ref{fig4}, with the corresponding parameters listed in
Table~\ref{tab2}. The SBPL model is generally better than the
Schechter function, with a correlation coefficient in the range of
$R^{2} \approx 0.98$ -- 0.993, compared to $R^{2} \approx 0.93$ --
0.986 for the Schechter model. Note that the sample size of the
rising group is relatively small, so the corresponding fits are
less reliable and the model parameters are not well constrained.
Despite minor fluctuations in the original data points, the smooth
best-fit functions adequately capture the overall trends relevant
for comparison. In short, we see that the luminosity functions are
largely similar for the three subsamples, supporting a common
origin for plateau GRBs that is basically independent of the
observed plateau slope.

\subsection{Event Rate}

\begin{figure*}[ht!]
\centering
\includegraphics[width=0.55\textwidth, trim=0 0 0 0, clip]{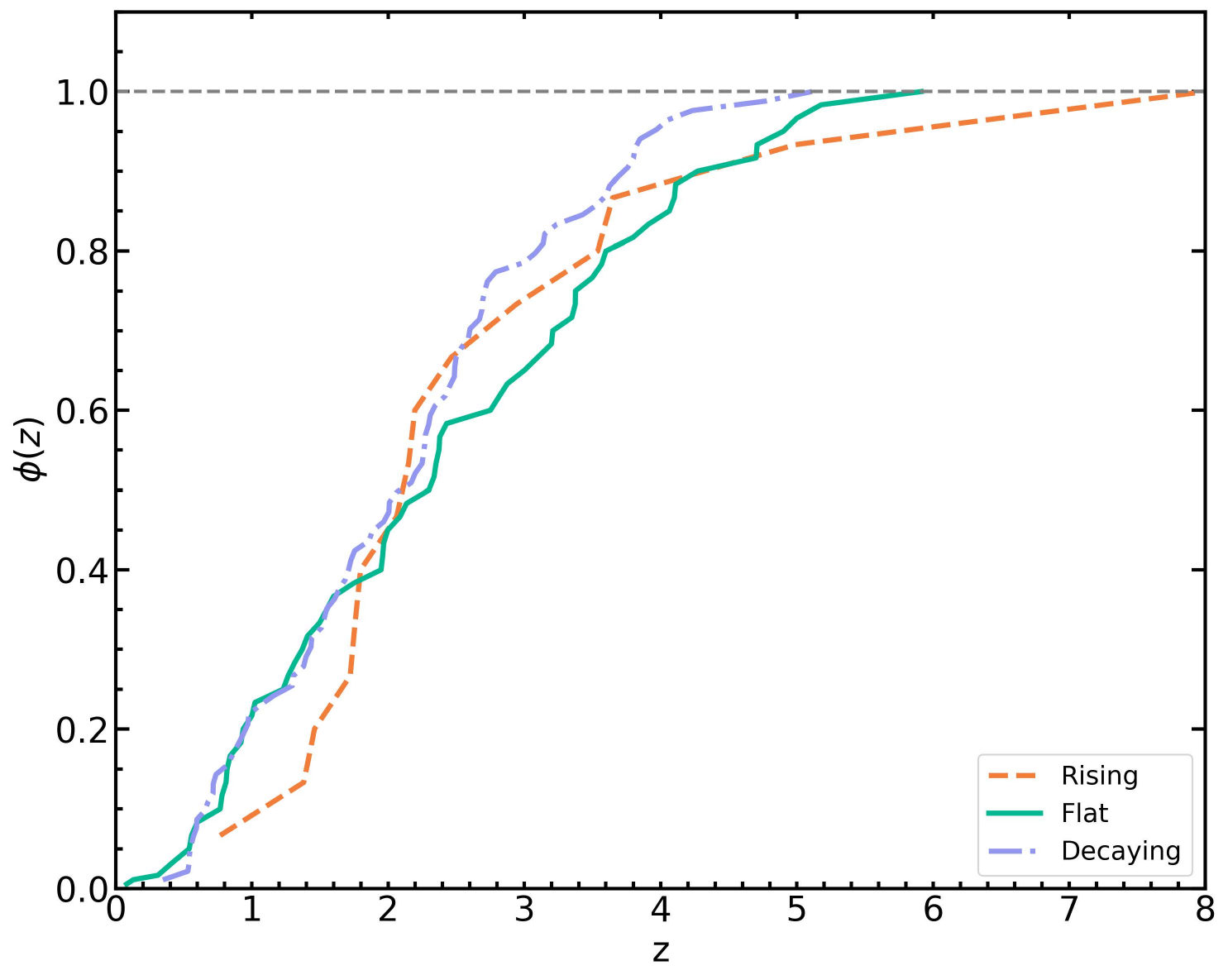}
\caption{ Normalized cumulative redshift distributions of the
three subclasses. The dashed, solid, and dash-dotted lines
represent the rising, flat, and decaying plateau groups,
respectively. } \label{fig5}
\end{figure*}

Figure~\ref{fig5} shows the normalized cumulative redshift
distributions of the three subsamples, calculated using
Equation~(\ref{redshiftFunction}). Nearly all GRBs are in the
range of $0 < z \lesssim 6$. The only outlier is GRB~090423
($z\approx8$; \citealt{2009Natur.461.1254T, 2009Natur.461.1258S}),
which belongs to the rising-plateau subsample. Although its
rest-frame duration is short ($T_{\rm 90,rest}\sim1.1$s), it is
widely believed to be a collapsar-type long GRB when its
observational features are synthesized.

\begin{figure*}[ht!]
\centering
\includegraphics[width=0.97\textwidth, trim=0 0 0 0, clip]{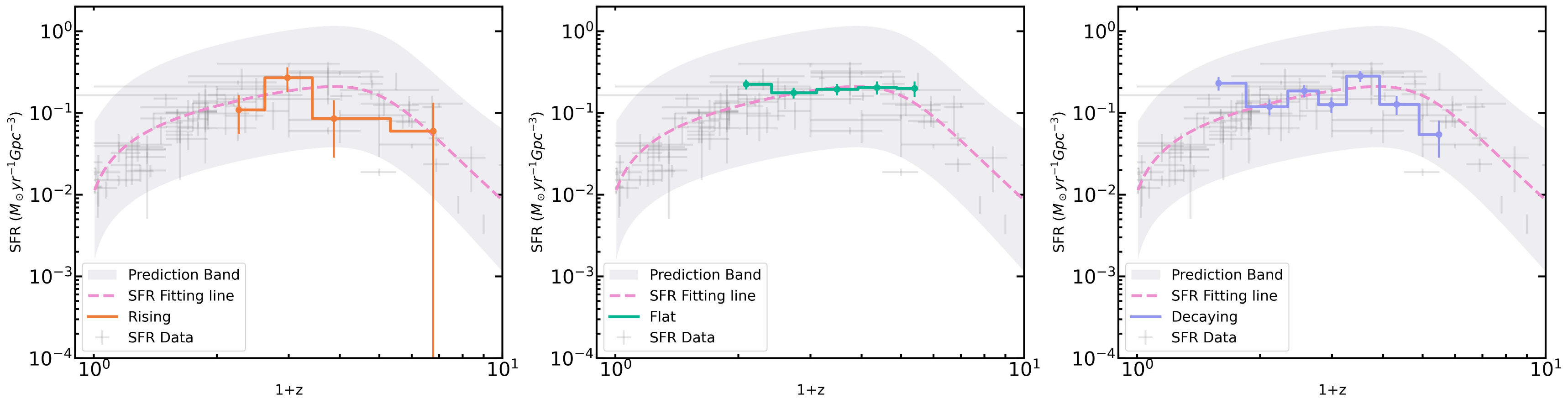}
\caption{ Event rates of the rising, flat, and decaying plateau
subsamples. Step lines show the inferred event rates; scatter data
points show observed SFR. The dashed lines show the best-fit SFR
curve and the gray shaded region marks its 99\% confidence level,
adopted from \cite{2022MNRAS.513.1078D}. } \label{fig6}
\end{figure*}

Figure~\ref{fig6} illustrates the event rate of the three
subsamples, computed by using Equation~(\ref{formationrate}). The
event rate is scaled by an arbitrary factor for direct comparison
with the SFR. The three groups exhibit similar redshift evolution:
the event rate is weakly dependent on the redshift, and is largely
consistent with the observed SFR data in the redshift range of 1
-- 5. Such a close similarity among the rising, flat, and decaying
groups suggests that the X-ray plateau emission is most likely
governed by a common mechanism. Otherwise, even if multiple
physical channels contribute to the plateau formation, their
relative contributions should not vary significantly across the
three categories. Additionally, the agreement with the cosmic SFR
further suggests that GRBs exhibiting a plateau phase are
predominantly residing in active star-forming environments,
supporting their connection with the death of massive stars.

\section{Slope Criteria and Definition of the Subsamples}
\label{sec:mctest}

In our study, we have divided GRBs with X-ray plateaus into
\textit{Rising} ($-0.5<\alpha_{1}\leq-0.1$), \textit{Flat}
($-0.1<\alpha_{1}\leq0.1$), and \textit{Decaying}
($0.1<\alpha_{1}\leq0.5$) groups. Across these groups, it is found
that the luminosity functions, cumulative redshift distributions,
and event rates exhibit similar shapes. The similarity suggests
that the plateau GRBs with different plateau slopes may have a
common origin.

In the previous analysis, the intermediate group (i.e. the \textit{Flat}
plateau subsample) is defined by a slope criteria of 0.1. That is,
the flat group meet the requirement of $-0.1<\alpha_{1}\leq0.1$.
One concern is whether the result is dependent on the choice of
the criteria. To test this, we performed a Monte Carlo analysis in
which the threshold of $|\alpha_1|$ (originally 0.1) varies
uniformly between 0.05 and 0.15. The range of 0.05 -- 0.15 was
chosen to ensure that each group contains enough GRBs for
meaningful statistics, particularly for the rising group. For each
of the randomly evaluated threshold, the sample was reclassified
and the luminosity functions, redshift distributions, and event
rates of the subsamples were recomputed. We repeated the process
10,000 times, which yielded 10,000 realizations of the key
features for the three groups, allowing for a thorough assessment
of the robustness of our results.

\begin{figure*}[ht!]
\centering
\includegraphics[width=0.95\textwidth, trim=0 0 0 0, clip]{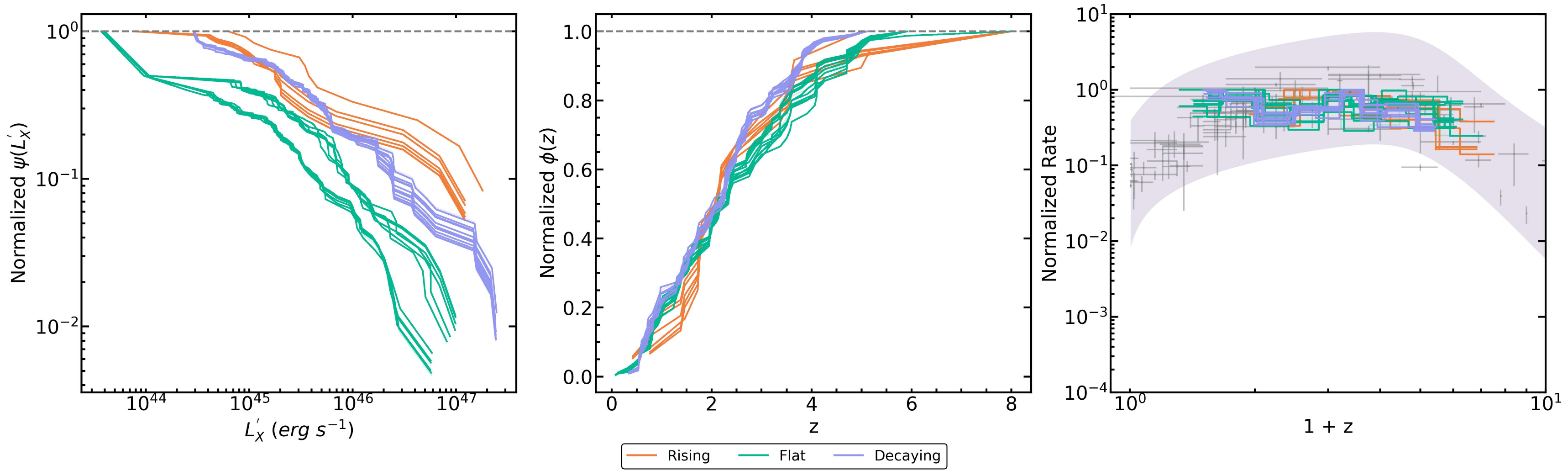}
\caption{Results of 10,000 Monte Carlo simulations for the slope
criteria.  The left, middle, and right panels show the normalized
luminosity functions, cumulative redshift distributions and event
rates, respectively. The orange, green, and purple lines
correspond to the rising-, flat-, and decaying-plateau subsamples.
 } \label{fig7}
\end{figure*}

Figure~\ref{fig7} shows all Monte Carlo realizations, normalized
for comparison. We see that for each subsample, the realizations
cluster tightly, indicating that the luminosity function, redshift
distribution, and event rate are insensitive to the precise choice
of the slope criteria. The three subsamples also remain highly
similar across all realizations. In particular, their normalized
event rates are highly clustered and overlapped, which are quite
flat for $z\lesssim 5$, consistent with the cosmic SFR. Our
simulations demonstrate that the similarity among the three
subsamples groups is robust and is not an artifact of the adopted
boundaries of $\alpha_{1}$. Thus, GRBs with X-ray plateaus are
governed by a common underlying mechanism although the plateau
slope varies markedly.

\section{Conclusion}
\label{sec:conclusion}

The origin of the X-ray plateau in GRB afterglows remains
uncertain, with numerous models proposed and no consensus yet
established. Motivated by the possibility that the timing index
$\alpha_{1}$ of the plateau may encode information about the
underlying mechanism, we examined whether different values of
$\alpha_{1}$ correspond to different plateau populations.

A large sample of 185 \textit{Swift}/XRT GRBs with a well-defined
plateau is compiled in this study. They are further divided into
rising, flat, and decaying subsamples using a threshold of
$\pm0.1$. The luminosity function, cumulative redshift
distribution and event rate are derived for each group. It is
found that these three functions exhibit similar behavior across
the subsamples, with no statistically significant differences. A
Monte Carlo test in which the slope criteria varies over a broad
range further confirms that the similarity is robust and is not an
artifact of the particular choice of the slope criteria.

Our results indicate that the plateau decay index alone does not
delineate distinct physical subclasses. Instead, GRBs with a
rising, flat, or decaying X-ray plateau exhibit statistically
indistinguishable properties, strongly pointing to a unified
mechanism for the plateau rather than drawing from different
progenitors or environments.

\section{Discussion}
\label{sec:Discussion}

Our population-level analysis shows no distinct separation among
rising, flat, and decaying subsamples. The shallow-decay index
exhibits a continuous and uni-modal distribution, consistent with
previous studies involving plateau luminosity, isotropic energy,
break time, and Dainotti-relation residuals
\citep{2010ApJ...722L.215D, 2019ApJS..245....1T,
2022ApJ...924...69Y}. These results suggest that X-ray plateaus
form a single, continuous population rather than discrete
subtypes.

\begin{figure*}[ht!]
\centering
\includegraphics[width=0.5\textwidth, trim=0 0 0 0, clip]{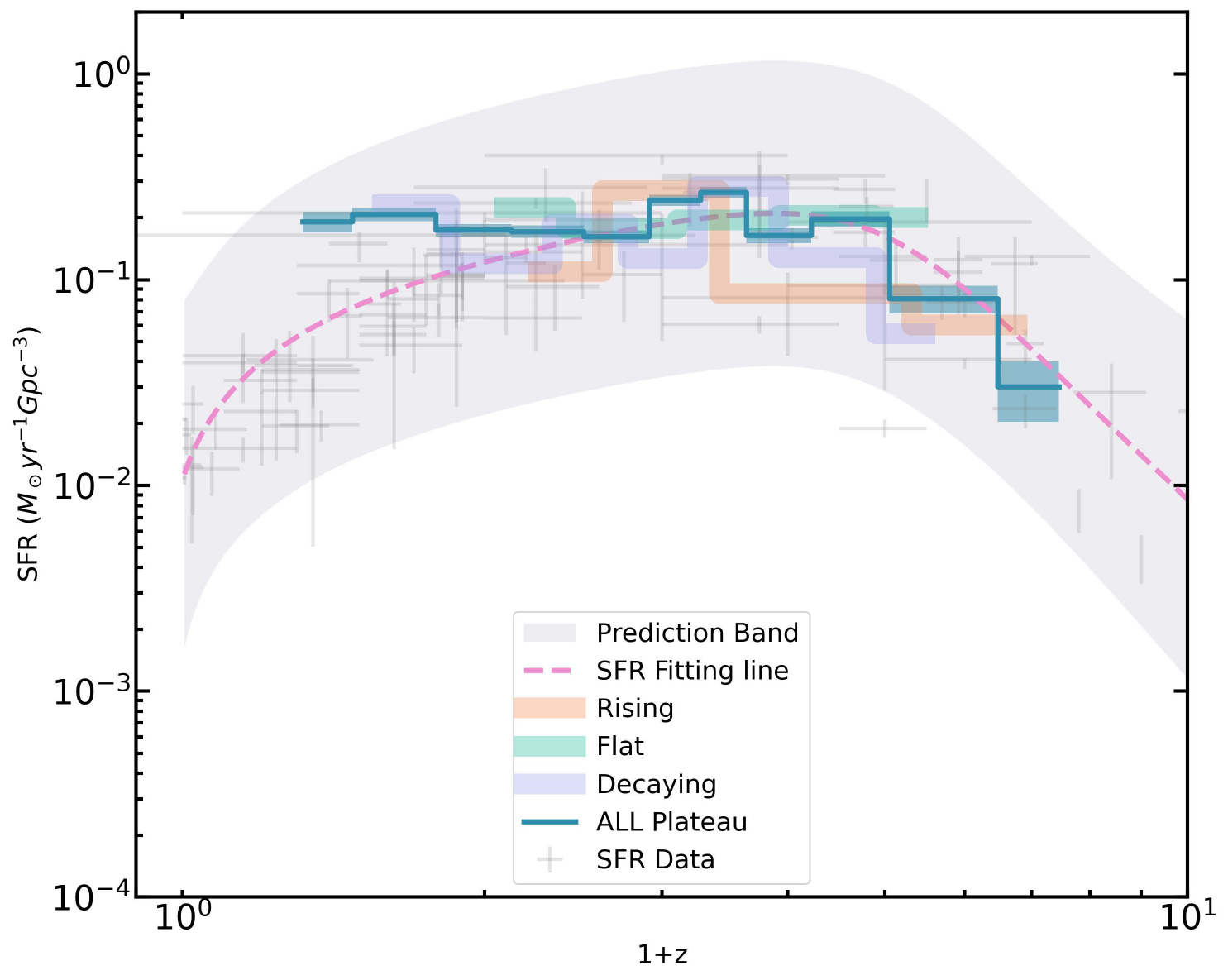}
\caption{ Event rates derived from the full sample of 185 X-ray
plateau GRBs. The solid step line shows the event rate of the full
plateau sample, while the shaded step lines represent the event
rates of the rising, flat, and decaying groups, respectively. }
\label{fig8}
\end{figure*}

We have also taken all the 185 X-ray plateau GRBs as a single
sample and recomputed the event rate. The result is plotted in
Figure~\ref{fig8}. We see that the event rate is again independent
on the redshift as long as $z<4$, which is consistent with that of
the rising, flat, and decaying subsamples. It generally follows
the SFR, with only a mild low-redshift deviation, similar to the
feature reported by \cite{2025ApJ...990...69K}, who interpreted
the deviation as a consequence of GRBs being biased tracers of
star formation. Note that comparing with the study of
\cite{2025ApJ...990...69K}, a more stringent plateau-slope
criterion of $|\alpha_1| \le 0.5$ was applied in our sample.

\cite{2024ApJ...960...77D} recently analyzed 310 \textit{Swift}
GRB X-ray afterglows and divided them into four temporal
categories based on the presence of flares, plateaus, and breaks.
According to their classification, the Category I GRBs are
featured by an X-ray flare along with two breaks; Category II GRBs
have two breaks and a plateau; Category III GRBs have two breaks
but without a plateau; Category IV GRBs have no breaks and no
plateaus. They found that both the $\gamma$-ray hardness ratio and
the peak energy of bursts of types I, II, and III are identically
distributed and are significantly lower than the corresponding
parameters of type IV bursts in statistics. Our sample is somewhat
similar to their Category II. The difference is that we further
divide this sample into three subsamples by considering the
plateau slope.

Our population-level analysis indicates that the plateau slope
$\alpha_1$ does not delineate separate GRB populations at the
macro level. Instead, the difference in $\alpha_1$ likely reflects
micro-physical variations within a unified central-engine
framework. For instance, a continuous variation in the magnetar
spin period, surface magnetic field, or ellipticity naturally
produces a broad spectrum of plateau slopes
\citep{2010MNRAS.402..705L}. Event-level analyses also reveals
smooth, correlated variations among these parameters
\citep{2025ApJS..280...45L}, providing a physical basis for the
observed $\alpha_1$ continuity. In the future, a larger,
well-sampled plateau dataset could enable more in-depth
population-level constraints on the central-engine and
energy-injection efficiency.

Recently, several groups applied the Lynden--Bell's $C^{-}$ 
method to fast X-ray transients (FXTs) detected by the Einstein Probe 
to estimate their event rates, although the current samples are still 
limited \citep{2025ApJ...995L..53G, 2026ApJ...997L..15L}. 
It was found that the FXT event rate is comparable to that of 
typical long GRBs, suggesting a possible connection between the two 
populations. With the recent release of an expanded 
Einstein Probe FXT catalog \citep{Wu2025_inprep}, including 
sources exhibiting extended plateau-like evolution in their 
soft X-ray light curves, the growing uniform
sample may enable population-level analysis to constrain 
the origin of plateau-containing FXTs.

\section{Acknowledgements}
This study was supported by the National Natural Science
Foundation of China (Grant No. 12233002, 12273113), 
by the National Key R\&D Program of China (2021YFA0718500), 
by the China Postdoctoral Science Foundation (No. 2025M783225), 
by the Shandong Provincial Natural Science Foundation (No. ZR2023MA049),  
and by the Postgraduate Research \& Practice Innovation Program 
of Jiangsu Province (No. KYCX25\_0197). 
Y.F.H. acknowledges the support from the Xinjiang
Tianchi Program. J.J.G. acknowledges support from the Youth Innovation Promotion Association (2023331).\vspace{5mm}

\bibliography{main}{}
\bibliographystyle{aasjournal}

\end{document}